\documentclass[12pt]{article}

\usepackage{scicite}
\usepackage{epsfig}
\usepackage{times}
\usepackage{amsmath}
\usepackage{caption}
\usepackage{amssymb}

\newcommand{\be}{\begin{equation}}
\newcommand{\ee}{\end{equation}}
\newcommand{\bea}{\begin{eqnarray}}
\newcommand{\eea}{\end{eqnarray}}

\renewcommand{\epsilon}{\varepsilon}

\newenvironment{sciabstract}{%
\begin{quote} \bf}
{\end{quote}}

\newcounter{lastnote}
\newenvironment{scilastnote}{%
\setcounter{lastnote}{\value{enumiv}}%
\addtocounter{lastnote}{+1}%
\begin{list}%
{\arabic{lastnote}.}
{\setlength{\leftmargin}{.22in}}
{\setlength{\labelsep}{.5em}}}
{\end{list}}

\title{Experimental Spin Ratchet}

\author
 {Marius V. Costache$^1$ and Sergio O. Valenzuela$^{1,2,3*}$ \\
\\
 \normalsize{$^1$ Institut Catal\`{a} de Nanotecnologia (ICN),}\\
 \normalsize{Centre d'Investigaci\'{o} en Nanoci\`{e}ncia i Nanotecnologia (CIN2)}\\
 \normalsize{Campus UAB, Bellaterra, Barcelona E-08913, Spain}\\
 \normalsize{$^2$ Instituci\'{o} Catalana de Recerca i Estudis Avan\c{c}ats (ICREA), Barcelona E-08010, Spain}\\
 \normalsize{$^3$ Department of Physics, Harvard University, Cambridge, Massachusetts 02138, USA}\\
 \normalsize{$^\ast$To whom correspondence should be addressed; E-mail: SOV@icrea.cat } }

\date{}

\begin{document}

\baselineskip24pt

\maketitle

\begin{sciabstract}
Spintronics relies on the ability to transport and utilize the spin properties of an electron rather than its charge. We describe a spin rachet at the single-electron level that produces spin currents with no net bias or charge transport. Our device is based on the ground state energetics of a single electron transistor comprising a superconducting island connected to normal leads via tunnel barriers with different resistances that break spatial symmetry. We demonstrate spin transport and quantify the spin ratchet efficiency using ferromagnetic leads with known spin polarization. Our results are modeled theoretically and provide a robust route to the generation and manipulation of pure spin currents.
\end{sciabstract}

\vspace{20mm}

\textit{One-sentence summary}:
We propose and experimentally demonstrate a spin rachet at the single-electron level that generates pure spin currents.

\newpage

Brownian motors or ratchets refer to directed transport in the presence of a signal or perturbation that drives the system without an obvious bias in any preferred direction of motion. The perturbation generates useful work, for instance the transport of particles, when combined with asymmetry, often realized by a so-called ratchet potential (Fig. 1A) \cite{hanggi2005,reimann2002,hanggi2009}. Experimental realizations of ratchets are spread over many different fields of biology, chemistry and physics where the perturbation may be external to the system (e.g. induced by an experimentalist) or intrinsic to it (e.g. non-thermal noise). In mesoscopic structures, experiments have demonstrated ratchets in both the quantum and classical limits \cite{song1998,linke1999,villegas2003}. On such small scales, noise rectification with ratchets can be used to control particle transport and has become one of the most promising techniques for powering nanodevices \cite{hanggi2009}.

Because of the growing interest in the spin degree of freedom as a carrier of information \cite{zutic2004} as well as a means to address fundamental properties of quantum mechanics and quantum computation \cite{awschalom2002}, a variety of ratchets have been proposed in pursuit of unidirectional spin currents and spin control \cite{scheid2007,smirnov2008,flatte2008,scheid2009}. A pure spin ratchet \cite{flatte2008} generalizes the particle ratchet mechanism \cite{hanggi2005,reimann2002,hanggi2009}, enabling pure spin currents by means of broken spatial symmetry \cite{scheid2007,smirnov2008,flatte2008,scheid2009}. Thus, an indispensable hallmark for a spin ratchet is the breaking of the inversion symmetry for spin but not charge \cite{flatte2008}, whereby the ratchet-potential easy direction for one spin orientation is opposite to the ratchet-potential easy direction for the other spin orientation (Fig. 1A). Recent theoretical efforts employ mesoscopic semiconductors and non-uniform magnetic fields \cite{scheid2007}, asymmetric periodic structures with Rashba spin-orbit interaction \cite{smirnov2008}, and double-well structures combined with local external magnetic fields and resonant tunneling \cite{scheid2009}.

The concept of our spin-ratchet is different from what has been proposed before. A small-volume superconducting (S) island is connected via tunnel junctions with two normal metal electrodes [N(l) and N(r)] to form an asymmetric single electron transistor (SET) with different tunneling resistances (Fig. 1B). A voltage $V$ applied across the SET drives the system, whereas a voltage on the backgate $V_g$ sets the induced gate charge $Q = V_gC_g$ on the island, with $C_g$ the capacitive coupling between the island and the gate.

\begin{figure}[t]
\begin{center}\epsfig{file=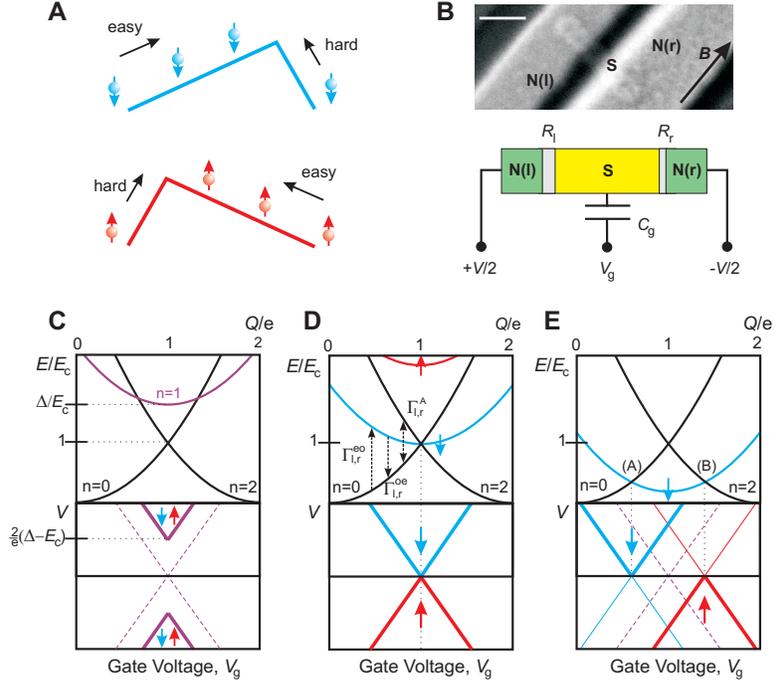,width=4 in}\end{center}\vspace{-5mm}
\caption{\footnotesize{Single electron transistor (SET) spin ratchet. \textbf{A}, In the presence of a ratchet potential and a driving force without a preferential direction, spin-down and spin-up electrons can be forced to move in opposite directions, giving rise to a spin current. \textbf{B}, Electron scanning microscope image of an SET spin ratchet. A small volume superconducting (S) island is contacted with two metal electrodes [N(l) and N(r)] via tunnel junctions with different tunnel resistances, $R_l>R_r$. The bar is 100 nm long. A voltage $V$ is applied across the electrodes, and a voltage $V_g$ on the backgate. \textbf{C-E}, SET energetics of Cooper-pair and quasiparticle states (top) and associated below-gap voltage thresholds (bottom) for single and two-electron transport at low temperatures for $B=0$ (\textbf{C}), $B=B_{SR}$ (\textbf{D}) and $B>B_{SR}$ (\textbf{E}). Dashed and solid lines represent the positions of the Andreev and quasiparticle conductance thresholds, respectively.}}\label{fig1}
\end{figure}

At low temperatures, parity effects in the superconducting island are important \cite{averin1992a,tuominen1992,eiles1993,hekking1993a}. When the number of conduction electrons $n$ is odd, there is necessarily one unpaired electron that is manifest as a quasiparticle excitation \cite{averin1992a,schon1998}. The ground state energy of the system for odd $n$ is higher than for even $n$ by the superconducting gap $\Delta$, which in our design is larger than the charging energy, $E_{\mathrm{c}}$ (Fig. 1C). In order to break the symmetry between spin-up and spin-down transport, a magnetic field $B$ is applied in-plane along the axis of the electrodes [spin up (down) refers to spins parallel (antiparallel) to $B$]. This field splits the quasiparticle levels (e.g. $n=1^{\downarrow}$ and $n=1^{\uparrow}$) by the Zeeman energy $E_Z=g \mu_B B$, where $g$ is the $g$-factor of the superconductor and $\mu_B$ the Bohr magneton, but it does not affect the Cooper-pair states (e.g. $n=0$ and $n=2$), which are singlet states, and it weakly reduces $\Delta$ because orbital-depairing is minimized by an in-plane $B$ \cite{ferguson2006} (Fig. 1, D and E). The $n=1^{\downarrow}$ state shifts down continuously with increasing $B$ and, at $B_{SR}=2(\Delta-E_c)/(g \mu_B)$ (Fig. 1D), it becomes degenerate with both the zero ($n=0$) and the one ($n=2$) excess Cooper-pair states for $Q/e =1$ ($e$ is the electron charge).

The spin ratchet effect occurs at $B=B_{SR}$. Insight into the underlying mechanism can be gained by analyzing the relevant charge transport processes and their occurrence rates. Single-electron tunneling processes in the $l$ and $r$ junctions cause transitions between even (e) $n=0,2$ and odd (o) $n=1^{\downarrow}$ states with rates $\Gamma_{l,r}^{oe}$ and $\Gamma_{l,r}^{eo}$, whereas two-electron Andreev processes cause transitions between even $n=0$ and $n=2$ states with rates $\Gamma_{l,r}^A$ (Fig. 1D). For a spin ratchet, the rate hierarchy $\Gamma_{l,r}^A \ll \Gamma_{l}^{oe} < \Gamma_{r}^{oe} \ll \Gamma_{l,r}^{eo}$ is required, where the $l$ junction transparency is chosen to be smaller than that of the $r$ junction. There, driving single-particle cycles (subsequent addition and removal of an electron from the SET island) results in a net spin current into one preferred direction in the following manner. A cycle that only uses transitions between $n=0$ and $n=1^{\downarrow}$ (cycle 01) only transports spin-down electrons through the SET, whereas a cycle that only uses transitions between $n=2$ and $n=1^{\downarrow}$ (cycle 21) only transports spin-up electrons. The essential ingredient to the spin ratchet mechanism is that, for $\Gamma_{l}^{oe} < \Gamma_{r}^{oe}$, cycle 01 dominates at positive $V$, while cycle 21 dominates at negative $V$. Hence, in both cases there is a net spin-up current from, say, left to right through the SET. Because the charge transferred is null in average when a voltage $V$ with zero mean is applied, the SET spin ratchet generates spin currents with no charge transport \cite{SOM}.

The thresholds for single and two-electron Andreev events in an SET fulfilling the above rate hierarchy are shown schematically in Fig. 1, C, D and E. At $B=0$, single electron transport sets in only for $V>2(\Delta-E_c)/e$, when the odd state is reached ($Q/e =1$). When $B$ is applied the Andreev and quasiparticle thresholds become closer and, at $B_{SR}$, they coincide. There, single electron transport is possible even at $V \sim 0$ and the spin ratchet is effective for an unbiased $V$, where the spin orientation of moving electrons changes sign at $V = 0$. For larger $B$, the ground state energetics of the SET fully separates cycles 01 and 21 around the degeneracy points (A) and (B) \cite{ferguson2006}. There, the asymmetric SET acts as a diode that resolves spin \cite{SOM}.

We have realized the proposed SET spin ratchet using electron-beam lithography and shadow evaporation techniques \cite{SOV2006}. The small (6 nm thick by 40 nm wide by 250 nm long) superconducting island is made from aluminium, which is oxidized and contacted with two metal leads. Sequential deposition of the leads from two different angles allowed us to generate distinct tunneling resistances in the junctions \cite{SOM}. We verified the spin-ratchet mechanism in Fig. 1 by means of ferromagnetic (F) leads made of CoFe that were used as spin detectors (FSF device). The spin polarization sign-change at $V=0$ is preserved, as when using normal leads, but the effective polarization of the leads, $P_F$, measures the relative contribution of cycles 01 and 21. For a quantitative measurement of the spin-ratchet efficiency, we independently determined $P_F$. We accomplished this using similarly fabricated junctions embedded in nonlocal spin devices for which we obtained $P_F\sim 0.28$ \cite{SOV2006,SOV2009}.

The electron transport properties of such an FSF SET were fully characterized by means of differential conductance d$I$/d$V$ measurements at above-gap voltage bias from which we estimated $\Gamma_l^{oe}\approx 8~10^6 ~\mathrm{s}^{-1} <$ $\Gamma_r^{oe}\approx 4~10^7 ~ \mathrm{s}^{-1} \ll$ $\Gamma_{l,r}^{eo}\approx 5~10^9 ~ \mathrm{s}^{-1}$ \cite{SOM}. Fig. 2 shows the evolution of d$I$/d$V$ as a function of the magnetic field at below-gap bias for this device. At $B=0$, we observe a symmetric response about $V=0$ (Fig. 2A). There, d$I$/d$V$ is zero within the sensitivity of our measurements for voltage magnitudes below the gap, except at the quasiparticle thresholds, where it presents a peak whose intensity is nearly independent of $V$ and $V_g$ \cite{Note1}. The below-gap quasiparticle thresholds cross at about $V_0=$ 259 $\mu$V (Fig. 2A). This is in agreement with $V_0 \sim 2(\Delta-E_c)/e$ (Fig. 1C) when using $E_c= 170$ $\mu$eV and $\Delta \approx 303$ $\mu$eV as obtained from the above-gap thresholds [Fig. S2 \cite{SOM}]. At $B=1$ T, $V_0$ decreases to 94 $\mu$V due to $E_Z$. At $B=1.5$ T, $V_0$ becomes zero and the SET is in the pure spin ratchet regime (Fig. 1D) \cite{Note2}.

\begin{figure}[h]
\vspace{2mm}
\begin{center}\epsfig{file=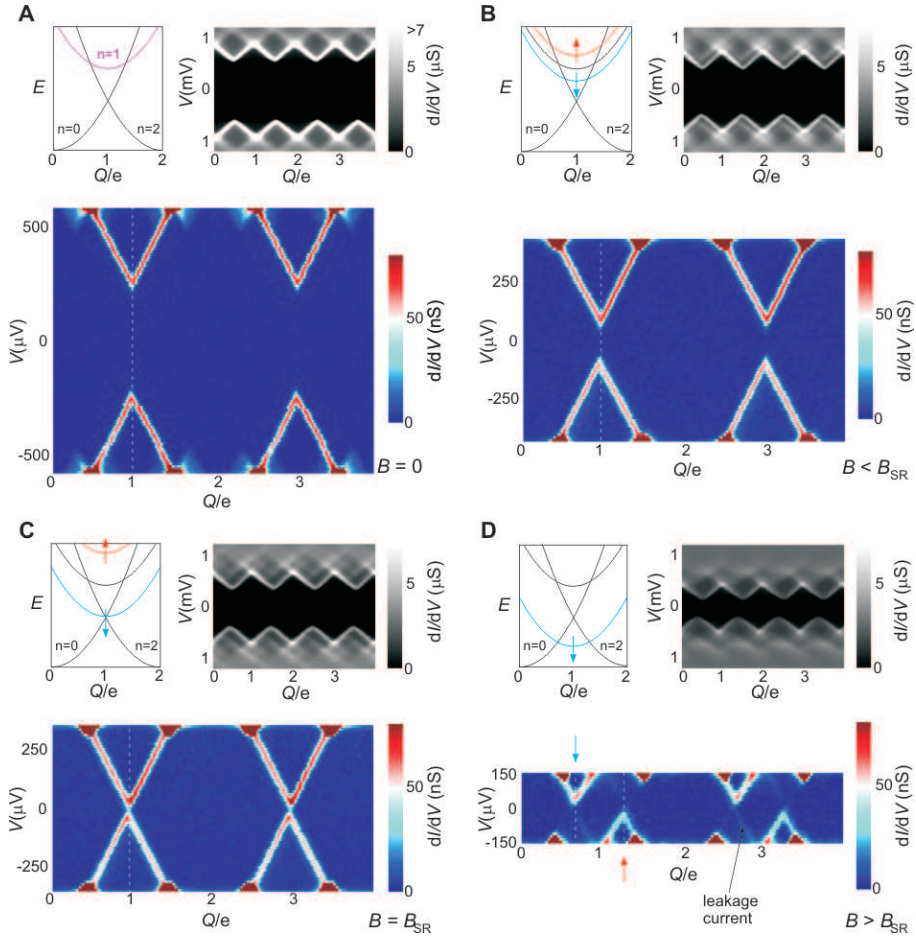,width=5 in} \end{center}
\caption{\footnotesize{Spin transport regimes in an applied magnetic field and characteristics of an SET spin ratchet. \textbf{A}, $B=0$. \textbf{B}, $B=1$ T. \textbf{C}, $B=1.5$ T. \textbf{D}, $B=2.5$ T. The top panels represent the SET energetics of Cooper-pair and quasiparticle states at the associated $B$ (left) and show the above-gap response d$I$/d$V$ versus gate $V_g$ and bias $V$ voltages (right). The bottom panels show the below-gap transport in the SET (black area in the above-gap d$I$/d$V$ plots).}}\label{fig2}
\end{figure}

Of key importance, the differential conductance at $B\neq0$ (Fig. 2, B and C) is no longer symmetric about $V=0$, presenting a larger magnitude for $V >0$ than for $V<0$ along the below-gap quasiparticle thresholds. This observation is consistent with the description in Fig. 1 and represents an experimental confirmation of the spin ratchet effect. Indeed, the asymmetry results from $P_F$ and the fact that the current across the SET for positive and negative $V$ has opposite spin polarization. The leads are always magnetized parallel to each other along the $B$ direction and, because $P_F > 0$, they favor the dominant spin-down current cycle 01 at $V>0$ and hinder the dominant spin-up current cycle 21 at $V<0$. We quantify such a transport asymmetry using the parameter $\beta = (G^+_{p}-G^-_{p})/(G^+_{p}+G^-_{p})$, where $G^+_{p}$ = d$I$/d$V\rfloor_{peak}$ ($V>0$) and $G^-_{p}$ = d$I$/d$V\rfloor_{peak}$ ($V<0$) are the values of the peak conductances along the dotted white lines in Fig. 2. At $B=0$ (Figs. 2A and 3A), $\beta$ is zero within the sensitivity of our measurements, as expected. At $B=1$ T and $B=1.5$ T (Fig. 2, B and C, and Fig. 3, B and C), the difference between $G^+_{p}$ and $G^-_{p}$ becomes apparent resulting in $\beta\sim 0.14$ in both cases.

\begin{figure}[t]
\begin{center}\epsfig{file=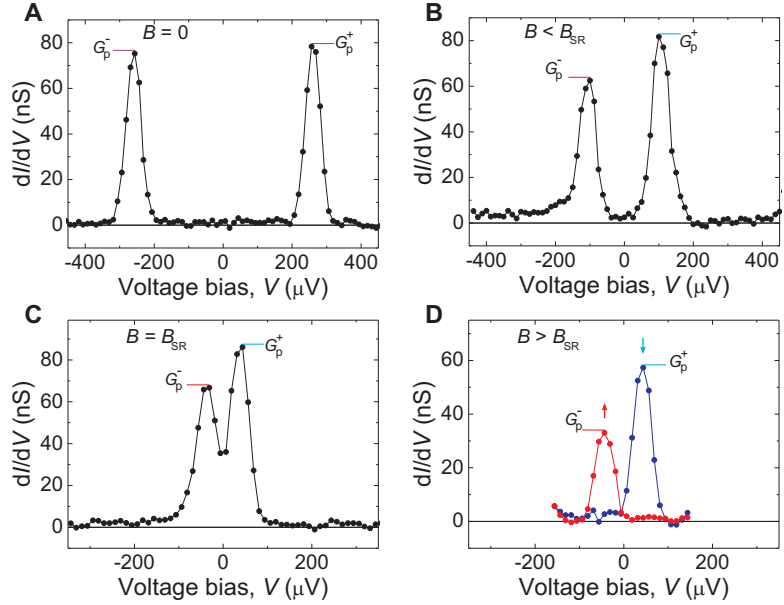,width=4.5in}\end{center}\vspace{-3mm}
\caption{\footnotesize{Spin filtering. \textbf{A}, $B=0$. \textbf{B}, $B=1$ T. \textbf{C}, $B=1.5$ T. \textbf{D}, $B=2.5$ T. Differential conductance d$I$/d$V$ versus $V$ cross-sections along the dotted white lines in Fig. 3. In \textbf{D}, the red and blue curves are cross-sections along the white lines indicated with red and blue arrows in Fig. 2D, respectively.}}\label{fig3}
\end{figure}

We define the spin-ratchet efficiency $\eta_{SET}$ as equal to the spin filtering capability $\eta_{SET}\approx(1-\alpha)/(1+\alpha)$ of our device, where the ratio $\alpha=\Gamma_{l}^{oe}/\Gamma_{r}^{oe}\approx R_r/R_l$ measures the asymmetry of the SET and $R_{l,r}$ are the associated normal tunnel resistance of junctions $l, r$ (Fig. 1B). For $\alpha \sim 0$, nearly perfect filtering, that is, $\eta_{SET}\sim 1$, is achieved. In such scenario, $\beta$ directly measures the effective polarization of the leads; that is, $\beta=P_F =0.28$. For $\alpha > 0$, a decrease in filtering efficiency is expected and therefore $\beta$ should decrease accordingly as $\beta \approx \eta_{SET} P_F$. For our device $R_l \approx 350$ k$\Omega$ and $R_r \approx 70$ k$\Omega$ and $\alpha \sim 0.2$. We thus estimate $\eta_{SET} \sim 0.67$ and $\beta \approx \eta_{SET} P_F \sim 0.19$, a value that is somewhat larger than that obtained with our measurements ($\beta\sim 0.14$), which results in $\eta_{SET} \approx 0.5$. This discrepancy could be related to the uncertainty in the estimation of $R_{l,r}$ or to Andreev reflections in one of the junctions, which could contribute an unpolarized component to the total current.

At magnetic fields $B > B_{SR}$, where the spin-up and spin-down quasiparticle thresholds are resolved, the SET behaves as a diode that filters spin-up or spin-down quasiparticles (Figs. 2D and 3D). Namely, the current should be fully spin-down polarized for $V_g$ about the degeneracy point (A) and spin-up polarized for $V_g$ about the degeneracy point (B) in Fig. 1E. Accordingly, we calculate $\beta$ from the conductance peaks along the two dotted lines in Fig. 2D obtaining $\beta\sim 0.26$, which is close to $P_F \sim 0.28$ and indicates a filtering efficiency larger than 0.9.

Lastly, we stress that the spin ratchet effect is related to quasiparticle tunneling through the high-transparency junction \cite{Note1}. To further show this, we fabricated devices with a normal (N) metal lead made of Cu connected to the low-transparency junction (NSF). Here, $R_l \approx 650$ k$\Omega$ and $R_r \approx 70$ k$\Omega$. As the high-transparency tunnel barrier connected to the ferromagnetic lead controls the transport, $\beta$ should remain close to $P_F$, when calculated as in Fig. 3D. Moreover, because $R_r$ in this device is estimated to be of the same order of magnitude as that of the FSF device, the conductance peaks should not be significantly affected. Both these observations agree with the experimental d$I$/d$V$ results shown in Fig. 4. At $B=0$ (Fig. 4A), $\beta$ is again zero within the sensitivity of our measurements and, at $B > B_{SR}$ (Fig. 4B), $\beta\sim 0.25 \sim P_F$, whereas the magnitudes of the conductance peaks compare well with those shown in Fig. 3.

\begin{figure}[h]
\begin{center}\epsfig{file=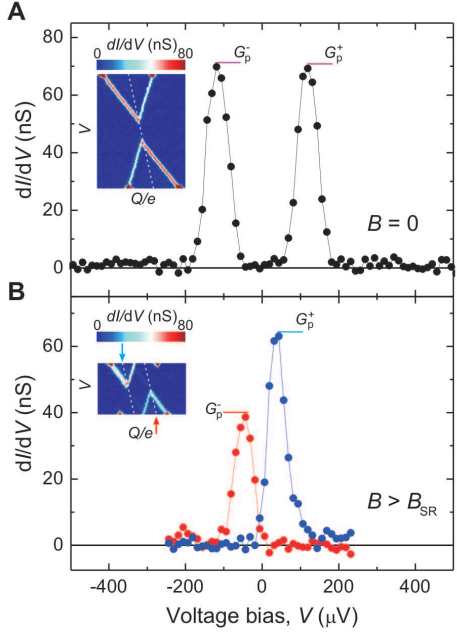,width=2.7in}\end{center}\vspace{-8mm}
\caption{\footnotesize{Spin filtering detection using an NSF sample. \textbf{A}, $B=0$. \textbf{B}, $B=2$ T. The insets show d$I$/d$V$ versus gate $V_g$ and bias $V$ voltages. The d$I$/d$V$ versus $V$ cross-sections (main panels) are taken along the corresponding dotted lines in the insets.}}\label{fig4}
\end{figure}
Spin ratchets represent a fundamentally new approach for spin current generation and detection, thus our research paves the way for a new means to study spin-related phenomena. Because the spin ratchets presented here work at the single-electron level, they can, for example, be used to initialize and readout the state of spin-based quantum bits \cite{awschalom2002} or to identify the spin orientation of single electrons in a test of the Einstein-Podolsky-Rosen paradox \cite{Einstein1935} with spin-entangled electrons \cite{bena2002,loss2003,hofstetter2009,cadden2009,herrmann2010}.

\bibliographystyle{Science}

\begin{scilastnote}
\item We gratefully acknowledge discussions with and support from M. Tinkham. We thank I. \u{Z}uti\`{c} and Y. Tserkovnyak for discussions and W. D. Oliver, P. Gambardella and A. Bachtold for a critical reading of the manuscript. This research was supported in part by the Spanish Ministerio de Ciencia e Innovaci\'{o}n (MAT2010-18065, FIS2009-06671-E). Samples were made at the Center for Nanoscale Systems (CNS), Harvard University.
\end{scilastnote}

\newpage

\section*{Materials and Methods}
Our SETs consist of a small-volume (6 nm thick by 40 nm wide by 250 nm long) aluminum (Al) superconducting island (S) connected to two nonsuperconducting electrodes, N(l) and N(r). Fig. S1 shows the main elements for their fabrication, which involve electron-beam lithography and multi-angle shadow evaporation to produce tunnel barriers \emph{in situ} as described in our previous work \textit{(S1, S2)}. A suspended shadow mask (Fig. S1A) is first created on a highly-doped Si $\langle100\rangle$ wafer with thermally grown oxide. To this end, we use a methyl-methacrylate (MMA)/poly(methyl-methacrylate) (PMMA) bilayer in combination with selective electron-beam exposure. The base resist (MMA) has a sensitivity that is $\sim$5 times larger than the top resist (PMMA), which allows us to generate a controlled undercut by exposing the bilayer with a dose that is sufficient to expose the MMA layer, but insufficient to expose the PMMA layer. The exposed bilayer is developed in an isopropanol / methyl-isobutyl-ketone solution and placed in a high-vacuum electron-beam evaporator (base pressure $<10^{-8}$ Torr).

The material evaporation sequence is shown in Fig. S1, B and C. First, we evaporate Al perpendicular to the substrate (yellow), which creates the superconducting island. Next, the Al is oxidized in pure oxygen ($100-150$ mTorr for 40 min) to generate insulating Al$_2$O$_3$ barriers. After the vacuum is recovered, the two electrodes, N(l) (blue) and N(r) (red), are sequentially deposited under angles of 50$^{\circ}$ relative to the substrate normal, where the substrate is tilted in opposite direction for  N(l) and N(r) (Fig. S1, B and D). The sequential deposition leads to different tunneling resistances $R_l$ and $R_r$; the difference between $R_l$ and $R_r$ can be enhanced by an additional oxidation step in between each lead deposition.

The three-angle metal deposition results in a threefold projection of all of the mask features with a spatial shift, except for the island, which is deposited only once. The axis of rotation [indicated by a dashed line in Fig. S1A] is selected such that the island feature at 50$^{\circ}$ tilting projects onto the side-wall of the top PMMA resist (Fig. S1D), and later on the deposited material is removed by lift-off.

Measurements were performed in a dilution refrigerator at 25 mK with a true four-point ac/dc data acquisition technique. A dc voltage and a small superimposed ac sine voltage (20 $\mu$V) are applied to the SET. Both the ac current component through the SET and the ac voltage across the normal leads are acquired using standard lock-in techniques. Therefore, the measurements both indicate true bias and conductance.

The differential conductance d$I$/d$V$ at above-gap voltages is used to determine the device parameters, including the junctions capacitances $C_l$ and $C_r$, the gate capacitance $C_g$, and the superconducting gap $\Delta$. From the d$I$/d$V$ thresholds in Fig. S2 (FSF sample), the following parameters are obtained: $\Delta = 303$ $\mu$eV, $C_l \sim C_r \approx 235$ aF, $C_g \approx$ 1.4 aF, $C_{\Sigma}= C_l + C_r + C_g \approx 470$ aF, and $E_c = e^2/2C_{\Sigma}\approx 170$ $\mu$eV. The resistances for the left and right junctions are estimated independently as $R_l=350$ k$\Omega$ and $R_r=70$ k$\Omega$ ($\alpha\sim 0.2$) from similarly fabricated isolated junctions and the total SET resistance $R_l + R_l$ = 420 k$\Omega$. Using these parameters, we estimate $\Gamma_l^{oe}\approx 8~10^6 \mathrm{s}^{-1} <$ $\Gamma_r^{oe}\approx 4~10^7 \mathrm{s}^{-1} \ll$ $\Gamma_{l,r}^{eo}\approx 5~10^9 \mathrm{s}^{-1}$. The effective polarization of the ferromagnetic leads, $P_F$, was obtained using similarly fabricated junctions embedded in nonlocal spin devices \textit{(S3-S5)} for which we obtained $P_F\sim 0.28$.

\section*{Supporting Text}

\paragraph{Tunneling Rates in a Single Electron Transistor Spin Ratchet.}
The spin ratchet proposed and experimentally demonstrated in the main text results from the specific occurrence rates of the relevant tunneling events in a single electron transistor (SET) comprising a superconducting island contacted to normal leads via different tunneling resistances. Zeeman splitting favors the trapping of a quasiparticle in the island with a specific spin orientation and makes the (spin-down) quasiparticle state ($n=1^{\downarrow}$) degenerate with both the zero ($n=0$) and the one ($n=2$) excess Cooper-pair states. As discussed below, quasiparticle tunneling onto or off the island is favored through the junction with the smallest tunnel resistance, mainly involving either the $n=0$ or the $n=2$ state in the transport, depending on the bias direction. This results in moving electrons with opposite spin orientation when the bias is reversed, the hallmark of the spin ratchet.

Fig. S3 shows the relevant charge transport processes and their corresponding rates for a single electron transistor in the spin ratchet regime (Fig. 1D, main text). The widths of the arrows represent the relative weight of the different rates. Fig. S3A concentrates on the relative rates magnitudes in general, whereas Fig. S3, B and C, focus on the effect of different tunneling resistances in the $l$ and $r$ junctions.

At low voltages and temperatures, only the states $n=0$, $n=2$ and $n=1^{\downarrow}$ are needed to describe the transport; low-probability cotunneling events to higher excited states \textit{(S6, S7) }can be disregarded, as we verified experimentally. Single electron tunneling processes in the $l$ and $r$  junctions cause transitions between even (e) $n=0,2$ and odd (o) $n=1^{\downarrow}$ states with rates $\Gamma_{l}^{oe,eo}$ and $\Gamma_{r}^{oe,eo}$, respectively, whereas two-electron Andreev processes cause transitions between even $n=0$ and $n=2$ states with rates $\Gamma_{l,r}^A$. As demonstrated in Refs. \textit{(S7, S8)}, odd-to-even transitions occur with a much smaller rate than even-to-odd transitions ($\Gamma_{l,r}^{oe} \ll \Gamma_{l,r}^{eo}$) because in the former a specific quasiparticle must be removed from the superconducting island whereas in the latter all of the quasiparticle states are involved (Fig. S3A). The rates $\Gamma_{l,r}^{oe}$ are usually known as escape rates and apply to tunneling events in which the single quasiparticle in the odd-state leaves the island but also to events in which an electron from a lead tunnels into the state paired with the existing quasiparticle \textit{(S7)}.

For an efficient spin ratchet, small $\Gamma_{l,r}^A$ are desirable because paired electrons do not contribute to the spin current. Based on this and the previous discussion, we consider $\Gamma_{l,r}^A \ll \Gamma_{l,r}^{oe} \ll \Gamma_{l,r}^{eo}$, a first condition that can be satisfied with proper device design as described in the next section. In this situation, the charge current in the SET is limited by the specific quasiparticle escape rates $\Gamma_{l,r}^{oe}$ and, when transitions to state $n=1^{\downarrow}$ become energetically favorable, the average occupation of $n=1^{\downarrow}$ is $\sim\Gamma_{l,r}^{eo}/(\Gamma_{l,r}^{eo}+\Gamma_{r,l}^{oe}) \sim 1$.

A key point for our proposed spin-ratchet mechanism is that the ground state energetics of the SET dictates that different junction transparencies result in transport of spins with opposite orientation for positive and negative $V$. Therefore, a second condition requires that $\Gamma_{l}^{oe} < \Gamma_{r}^{oe}$, where the $l$ junction transparency is arbitrarily chosen to be smaller than that of the $r$ junction. Fig. S3, B and C, show the rates that dominate the transport of the asymmetric SET when electrons flow from left to right and from right to left, respectively. Because $\Gamma_{l}^{oe} < \Gamma_{r}^{oe}$, a quasiparticle removal process is more likely associated with a tunneling event in which either a quasiparticle directly tunnels off the island to the right lead (Fig. S3B) or, for opposite bias, an electron from the right lead tunnels onto the island to form a Cooper-pair with an existing quasiparticle (Fig. S3C). Tunneling events through the low-transparency left-junction may occur but with smaller probability.
As a direct consequence, transport of electrons from left to right (Fig. S3B) mostly involves the $n=1^{\downarrow}$ and $n=0$ states (cycle 01) because cycling between the $n=1^{\downarrow}$ and $n=2$ requires an electron tunneling from the left lead to remove the quasiparticle. In an analogous way, transport of electrons from right to left (Fig. S3C) mostly involves the $n=1^{\downarrow}$ and $n=2$ states (cycle 21) because cycling between the $n=1^{\downarrow}$ and $n=0$ requires the quasiparticle to tunnel off the island to the left lead.

Note that the effective easy direction of motion for one spin is thus opposite to the easy direction of motion for the other spin, as required in a spin ratchet (Fig. 1A, main text). Cycle 01 results in a spin-down polarized current for left-to-right electron motion, whereas cycle 21 results in spin-up polarized currents for right-to-left electron motion and overall both cases contribute to a spin current in the same direction. The efficiency to generate this spin current is directly related to the parameter $\alpha=\Gamma_{l}^{oe}/\Gamma_{r}^{oe}$, which measures the asymmetry of the SET; the smaller $\alpha$, the more efficient is the spin ratchet. Because for opposite bias the rates involved are the same, the charge transferred is null in average when a voltage $V$ with zero mean (no net bias) is applied, thus the SET spin ratchet generates pure spin currents.

The spin-ratchet is realized at an applied magnetic field $B=B_{SR}$. For  $B>B_{SR}$, the asymmetric SET acts as a diode that resolves spin (Fig. 1E, main text). There, it is necessary to consider separately the degeneracies between $n=1^{\downarrow}$ and $n=0$ (A) and between $n=1^{\downarrow}$ and $n=2$ (B). In between the degeneracies, a single spin-down quasiparticle stays in the island. Around the first degeneracy point (A), only cycle 01 can be involved in transport: a spin-down quasiparticle may tunnel onto and off the island resulting in a spin-down current. Around the second degeneracy point (B), only cycle 21 can be involved in transport: a spin-up quasiparticle tunnels onto the island to form a Cooper-pair with the spin-down quasiparticle, and subsequently a spin-up quasiparticle tunnels off, breaking a pair and leaving a spin-down quasiparticle behind; a sequence that results in a spin-up current.

Note that, for finite $\alpha$, small spin-down and spin-up leakage currents at negative and positive $V$, respectively, are expected. Such currents are deduced from weak conductance peaks in the diode with reverse bias (Fig. 2D, main text). More efficient spin ratchets could be obtained in SETs designed with smaller $\alpha$, which could be achieved by incrementing the difference between $R_l$ and $R_r$. For $\alpha = 0.1$, the filtering efficiency $\eta_{SET}\approx(1-\alpha)/(1+\alpha)$ would exceed 0.8 and for $\alpha = 0.05$, it would exceed 0.9. Such values of $\alpha$, which require a small transparency in one of the junctions, could be achievable without a decrease in the overall current through the SET because transport is dominated by the tunneling rate $\Gamma_r^{oe}$ in the transparent junction $r$.

\paragraph{Device Design and Calculated Rates.}
The required rate hierarchy $\Gamma_{l,r}^A \ll \Gamma_{l}^{oe} < \Gamma_{r}^{oe} \ll \Gamma_{l,r}^{eo}$ is achieved by controlling the size and transparency of the tunnel junctions, and the superconducting island volume, $V_{S}$. First, one must note that although Andreev reflections depend on the precise geometry near the junctions as well as on impurities and scattering sites \textit{(S9-S12)}, they are second-order processes that are suppressed in junctions with low enough transparency. The rates $\Gamma_{l,r}^{oe}$ and $\Gamma_{l,r}^{eo}$, on the other hand, are first order processes that depend less dramatically on the junction transparencies, whereas $\Gamma_{l,r}^{oe}$ can be enhanced by reducing the sample volume due to the normalization of the wavefunction of an unpaired quasiparticle in the island \textit{(S7, S13)}.

Explicitly, $\Gamma_{l,r}^{eo} \sim (C_{\Sigma}R_{\Sigma})^{-1}$ \textit{(S13)}, where $C_{\Sigma}=C_l+C_r+C_g$ and $R_{\Sigma}=R_l+R_r$, with $R_{l,r}$ and $C_{l,r}$ the tunnel resistances and capacitances of junctions $l$ and $r$. In addition, the escape rate across junction $i$ ($i= l, r$) is given by \textit{(S7, S13)} $\Gamma_{i}^{oe}=(2e^2R_i\rho_n V_{S})^{-1}$, where $\rho_n$ is the normal density of states of the superconductor per unit volume (including spin). From this last relationship and the rate hierarchy, we obtain that the charge current beyond the quasiparticle thresholds $\sim e\Gamma_{i}^{oe}$ is independent of $V$ and that the spin ratchet efficiency is governed by $\alpha=\Gamma_{l}^{oe}/\Gamma_{r}^{oe}\approx R_r/R_l$.

Previous studies on NSN SETs have shown Andreev-reflection dominated transport at low temperatures when $\Delta > E_c$ \textit{(S14)}. There, given the fact that $\Gamma_{l,r}^A \gg \Gamma_{l,r}^{oe}$, an unpaired quasiparticle becomes effectively trapped in the island, thereby preventing any two-electron tunneling event and blocking the Andreev cycle. In those studies, the quasiparticle escape rates were at least an order of magnitude smaller than the dominant Andreev rates. However, this relationship is readily reverted, for instance, by decreasing the island volume in more than an order of magnitude while maintaining, or increasing, the junctions resistances. Specifically, in our devices the island volume is two orders of magnitude smaller, and the junctions resistances at least twice as large as those in Ref. \textit{(S14)}.

\newpage

\clearpage

\section*{Supporting Figures}

\vspace{35mm}

\begin{figure}[h]
 \begin{center} \epsfig{file=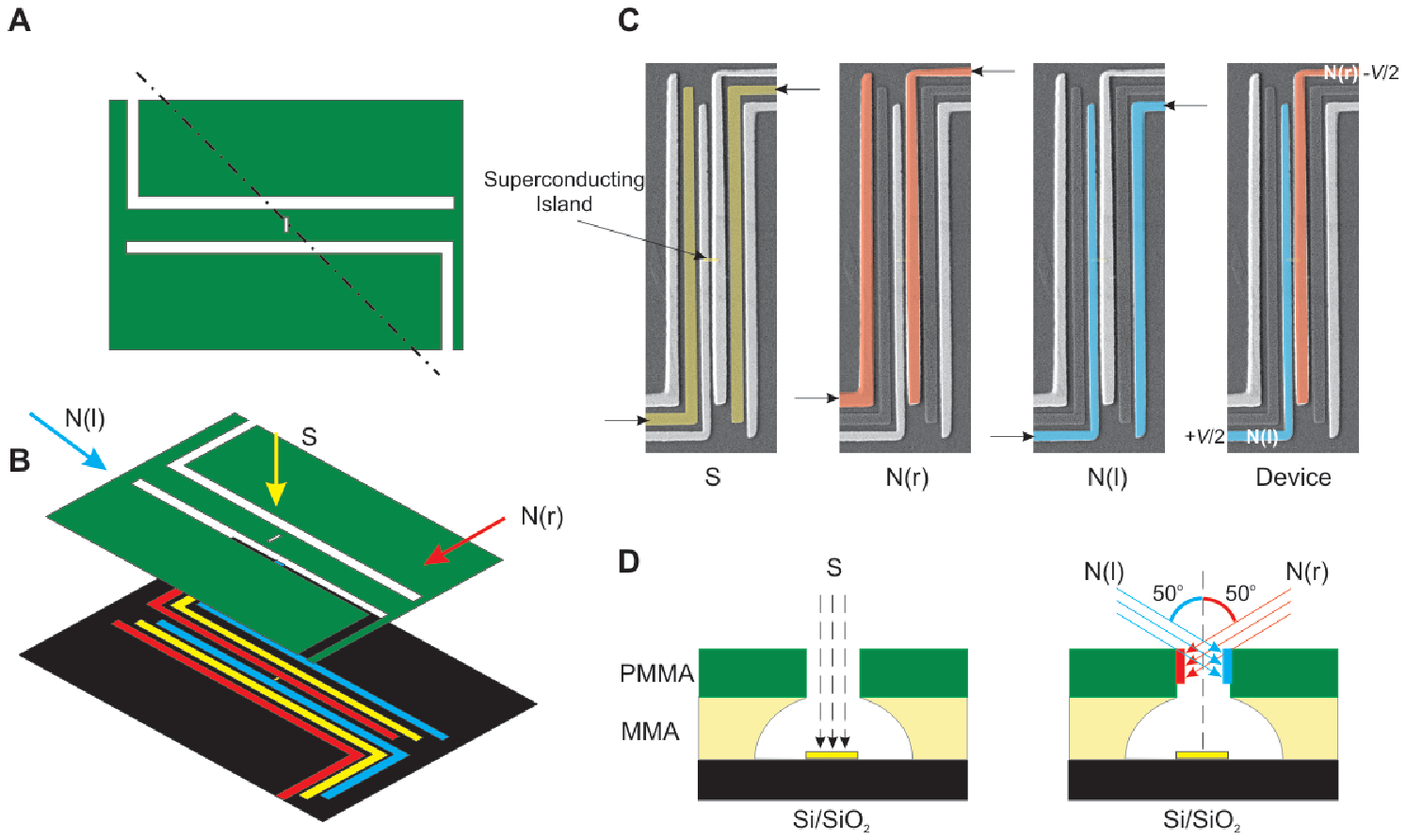,width=5.5in}\end{center}
\vspace{-5mm}\caption*{Figure S1:Sample fabrication. \textbf{A}, Design of the suspended MMA/PMMA mask for shadow evaporation. The dashed line represents the rotation axis for shadow evaporation. \textbf{B}, The device is fabricated by three sequential depositions as indicated by the arrows. Such a process results in a threefold projection of the mask. \textbf{C}, Scanning electron microscope images of a device showing, from left to right, the deposition sequence of the mask features. The deposited features in each step are indicated by superimposed colored areas and arrows. \textbf{D}, Vertical cross section of the mask. The projection of the island feature in the mask falls onto the side wall of the top resist, except for the Al evaporation, which is normal to the substrate.
}
 \label{figS2}
 \end{figure}

\begin{figure}
 \begin{center} \epsfig{file=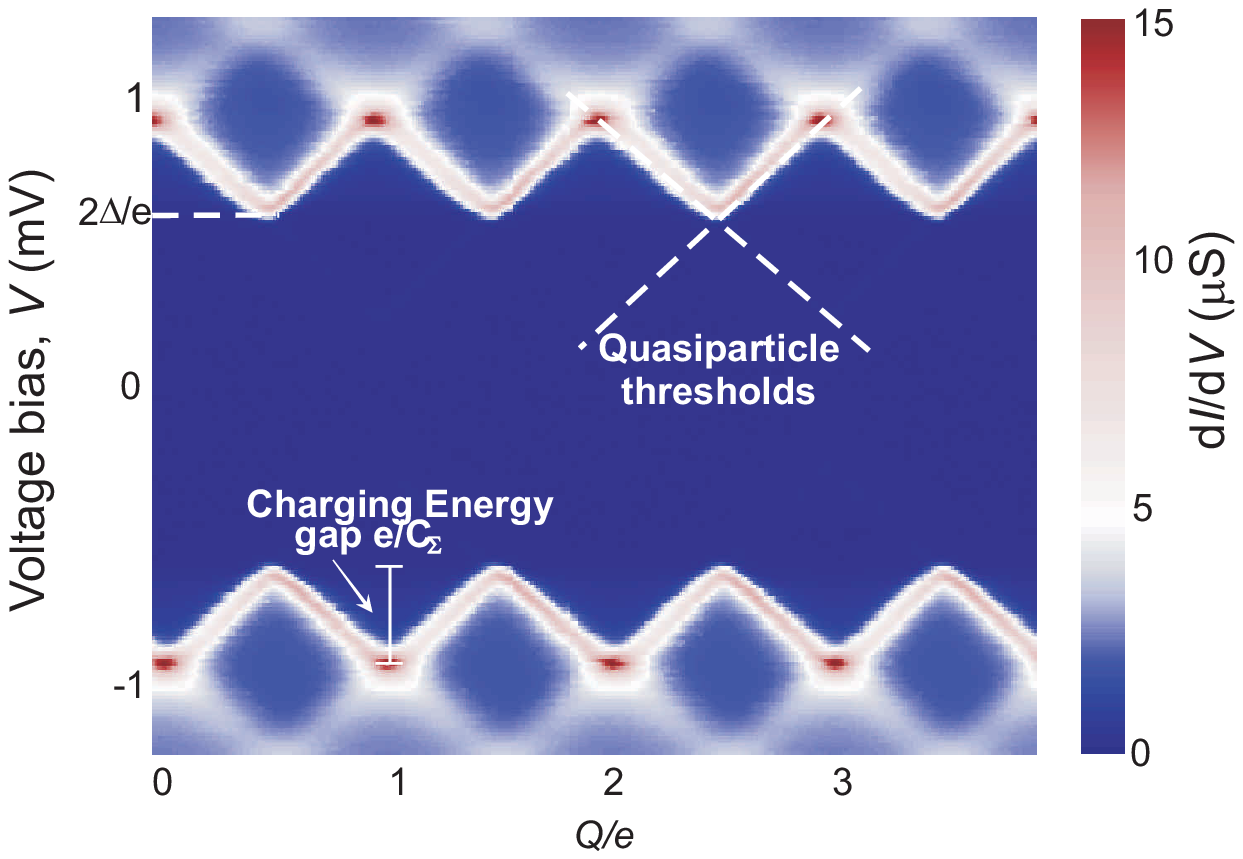,width=4in} \end{center}
\vspace{-5mm}\caption*{Figure S2: Experimental above-gap d$I$/d$V$ characteristics of a FSF device measured at 25 mK as a function of dc voltage $V$ across the SET and gate voltage $V_g$. The d$I$/d$V$ amplitude is represented by a color scale from blue (zero) to red (15 $\mu$S). From the voltage threshold for single quasiparticle events, the parameters $C_l \sim C_r \approx 235$ aF, $C_g \approx$ 1.4 aF, and $\Delta \approx 303$ $\mu$eV are obtained. The lines are guides to the eye for the threshold voltages above the gap. $B=0$, $T = 25$ mK.
}
 \label{figS3}
\end{figure}

\begin{figure}
 \begin{center}\epsfig{file=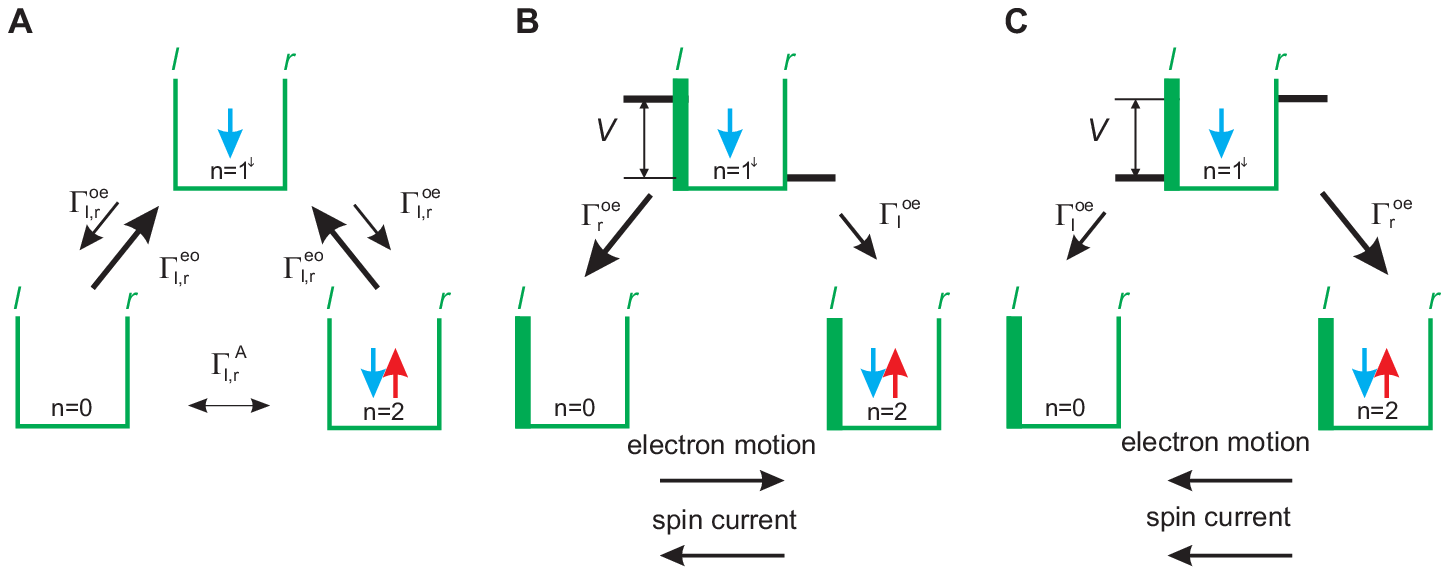,width=6in} \end{center}
\vspace{-5mm}\caption*{Figure S3: Illustration of the working principle of a single electron transistor (SET) spin ratchet; tunneling rates. \textbf{A}, Schematic representation of the allowed charge transport processes in the below-gap sequential-tunneling regime and the corresponding tunneling rates. Each green box depicts the SET in the indicated state ($n=0$, $n=2$, or $n=1^{\downarrow}$). Single-electron tunneling results in transitions between even (e), $n=0$ and $n=2$, and odd (o), $n=1^{\downarrow}$, states with rates $\Gamma_{l}^{oe,eo}$ and $\Gamma_{r}^{oe,eo}$. Two-electron Andreev processes cause transitions between even states with rates $\Gamma_{l,r}^A$. The SET is designed such that $\Gamma_{l,r}^A \ll \Gamma_{l,r}^{oe} \ll \Gamma_{l,r}^{eo}$. The arrows widths represent the relative magnitude of the rates. \textbf{B-C}, Dominant rates for positive and negative bias in the asymmetric SET at $B=B_{SR}$. The thickness of the left and right lateral walls of the green boxes represents the transparency of the tunnel junctions. The junction resistance to the left electrode is larger than that to the right electrode. For electrons moving towards the right (\textbf{B}), the electron current is spin-down polarized, whereas for electrons moving towards the left (\textbf{C}), the electron current is spin-up polarized. Overall, both processes contribute to a spin current with the same direction.
}
 \label{figS1}
\end{figure}

\end{document}